\documentclass[epj]{svjour}
\usepackage{graphics}
\begin{document}
\title{Microscopic membrane elasticity and interactions among membrane
  inclusions: interplay between the shape, dilation, tilt and
  tilt-difference modes}

\titlerunning{Microscopic Membrane
  Elasticity}
\author{Jean-Baptiste {\sc Fournier}
}                     
%
\mail{jbf@turner.pct.espci.fr}
\institute{Laboratoire de Physico-Chimie Th\'eorique,
  E.\,S.\,P.\,C.\,I.\,, 10 rue Vauquelin, 75231 Paris C\'edex 05,
  France.}
\date{Received: date / Revised version: date}
%
\abstract{A phenomenological Landau elasticity for the shape,
  dilation, and lipid-tilt of bilayer membranes is developed. The
  shape mode couples with the sum of the monolayers' tilt, while the
  dilation mode couples with the difference of the monolayers' tilts.
  Interactions among membrane inclusions within regular arrays are
  discussed. Inclusions modifying the membrane thickness and/or
  inducing a tilt-difference due to their convex or concave shape
  yield a dilation-induced attraction and a tilt-difference--induced
  repulsion. The resulting interaction can stabilize $2D$ crystal
  phases, with the possible coexistence of different lattice spacings
  when the dilation--tilt-difference coupling is large. Inclusions
  favoring crystals are those with either a long-convex or a
  short-concave hydrophobic core. Inclusions inducing a local membrane
  curvature due to their conical shape repel one another. At short
  inclusions separations, a tilt comparable with the inclusion's cone
  angle develops: it relaxes the membrane curvature and reduces the
  repulsion. At large separations the tilt vanishes, whatever the
  value of the shape-tilt coupling.
} 
\maketitle

\newcommand{\dd}{{\rm d}}
\newcommand{\pp}{\!<\!}
\newcommand{\pg}{\!>\!}
\newcommand{\eg}{\!=\!}

\section{Introduction}
\label{intro}
Bilayer membranes are formed by the self-assembly of amphiphilic
molecules in water of brine~\cite{isra}. The aliphatic chains of the
constituent molecules condense into an oily sheet that is shielded
from contact with water by the polar heads of the molecules.
Membranes can form various phases, e.g., lamellar ($L_\alpha)$,
vesicular ($L_4$), or sponge ($L_3$) phases~\cite{safran,luca}. Being
self-assembled systems with a conserved area, it is essentially the
competition between their curvature energy and their entropy that
determines their large scale behavior. In the standard macroscopic
theory, membranes are modeled as structureless surfaces with a
curvature elasticity~\cite{canhel1,canhel2}. This description, which has the
advantage to involve only two material constants, accounts for a large
number of universal properties and behaviors of amphiphilic
membranes~\cite{safran,luca,statsurf}.

On the other hand, a number of attempts have been made toward a more
{\em microscopic} description of membranes
\cite{marcel,owicki,huang,dan1,dan2,dan3}. The goal being to take into
account various structural parameters of the bilayer, such as its
thickness, the ordering of the chain segments, etc. As far as large
scale properties are concerned, these extra degrees of freedom are
irrelevant since they relax over microscopic lengths.  Nevertheless,
they can dictate important physical properties, such as adhesion
behaviors, the short-range interaction between membrane inclusions,
their aggregation properties, phase behaviors, etc. The aim of this
work is to construct an elastic model of membranes that connects
microscopic and macroscopic descriptions. Beside the standard shape
and dilation variables we shall consider as elastic variables the {\em
  tilts\/} of the lipids in both monolayers~\cite{seifert,euro}. The
model will be used to investigate the role of the monolayers tilts in
the interaction between membrane inclusions.

\section{Elastic model}
\label{sec:microel}

\begin{figure*}
\resizebox{1.0\textwidth}{!}
{\hspace{30pt}\includegraphics{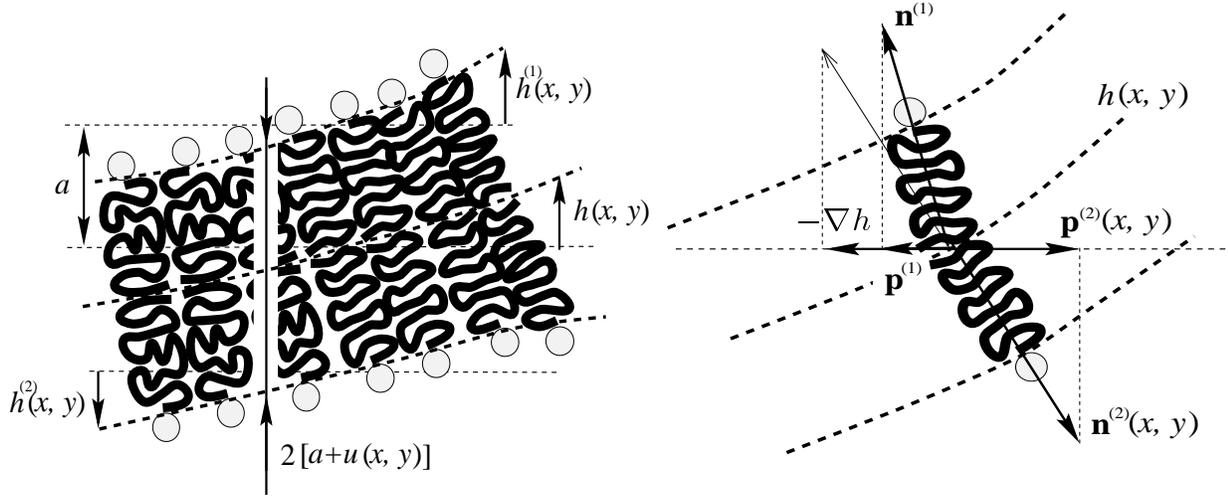}\hspace{30pt}}
\caption{a) Average membrane shape, $h$, and dilation, $u$, variables.
  b) Construction of the membrane average tilt ${\bf m}=\frac{1}{2}({\bf
    p}^{(1)}-{\bf p}^{(2)})+\nabla h$ and tilt-difference
  ${\bf\widehat m}=\frac{1}{2}({\bf p}^{(1)}+{\bf p}^{(2)})$ variables.}
\label{fig:def}
\end{figure*}

To construct an elastic model, one considers a distorsion free energy
depending on a particular set of structural parameters.  Implicitly,
this free energy results from the integration over all the microscopic
states compatible with these (fixed) parameters of the Boltzmann
weight associated with the microscopic Hamiltonian of the system. In
practice, based on the symmetries of the system, one writes an
expansion in powers of the structural parameters and their gradients.
The choice of the relevant parameters depends on which deformations
can be imposed externally on the system. We shall consider four
structural parameters: (1) the membrane thickness, which can be
modified, e.g., by the presence of an integral protein, (2) the
membrane average shape, in order to connect with the large-scale
theory and because it can be excited by a conically shaped inclusion,
and (3) the tilts of the molecules within each monolayer, that can be
independently excited by an inclusion with, e.g., a diamond-like
shape.

To simplify, we assume that the membrane undergoes only small
deviations with respect to its flat ground state. We denote by
$h^{(1)}(x,y)$ and $h^{(2)}(x,y)$ the vertical displacements (along
$z$) of the chain--water interfaces of the upper and lower monolayer,
respectively, with respect their positions in the flat unperturbed
state (Fig.~\ref{fig:def}a). For further use, let us define the
average {\em shape\/} $h(x,y)$ and the membrane {\em dilation\/} by
\begin{eqnarray}
  \label{def_h}
  h&=&{h^{(1)}+h^{(2)}\over2}\\
  \label{def_u}
  u&=&{h^{(1)}-h^{(2)}\over2}\,.
\end{eqnarray}
To construct the tilt variables, we introduce the vectors ${\bf
  p}^{(1)}(x,y)$ and ${\bf p}^{(2)}(x,y)$ defined in both monolayers
as the projections onto the $(x,y)$ plane of the unit vectors parallel
to the molecular direction and oriented from chain to polar head
(Fig.~\ref{fig:def}b). The tilts relative to the membrane normal are
measured by ${\bf m}^{(1)}={\bf p}^{(1)}+\nabla h$ and ${\bf
  m}^{(2)}={\bf p}^{(2)}-\nabla h$. Let us define the {\em average
  tilt\/} ${\bf m}(x,y)$ and the {\em tilt-difference\/} ${\bf\widehat
  m}(x,y)$ by
\begin{eqnarray}
  \label{def_p}
  {\bf m}&=&{{\bf m}^{(1)}-{\bf m}^{(2)}\over2}\,,\\
  \label{def_pchap}
  {\bf\widehat m}&=&{{\bf m}^{(1)}+
    {\bf m}^{(2)}\over2}\,.
\end{eqnarray}

\subsection{Shape and dilation distortion energy}

We start by constructing the most general quadratic free energy
expansion in powers of $h^{(1)}$ and $h^{(2)}$ and its first and
second gradient,
\begin{equation}
  h^{(\alpha)}\,;\quad h_{,i}^{(\alpha)}\,;\quad
  h_{,ij}^{(\alpha)}\,.
\end{equation}
Here, $\alpha=1,2$ is the monolayer label, and the
comma denotes partial derivation with respect to the coordinates $x$
and $y$. We write the free energy as $F=F^{(1)}+F^{(2)}+F^{(12)}$ with
all the interaction terms coupling $h^{(1)}$ and $h^{(2)}$ in
$F^{(12)}$. The symmetry of the bilayer imposes invariance with
respect to the transformation:
\begin{equation}\label{sym}
  h^{(1)}\to-h^{(2)},\quad h^{(2)}\to-h^{(1)}\,.
\end{equation}
Therefore, the most general quadratic form for $F^{(\alpha)}$ is
\begin{eqnarray}
  F^{(\alpha)}&=&
  (-1)^{\alpha}a_1\,h^{(\alpha)}+
  (-1)^{\alpha}a_2\,h^{(\alpha)}_{,ii}+
  a_3\,h^{(\alpha)2}\nonumber\\&&+
  a_4\,h^{(\alpha)}h^{(\alpha)}_{,ii}+
  a_5\,h^{(\alpha)}_{,i}\,h^{(\alpha)}_{,i}+
  a_6\,h^{(\alpha)2}_{,ii}\nonumber\\&&+
  a_7\,h^{(\alpha)}_{,ij}h^{(\alpha)}_{,ij}\,,
\end{eqnarray}
summation over repeated indices being understood. The interaction
energy, containing all the bilinear scalars, has the form
\begin{eqnarray}
  F^{(12)}&=&
  b_1\,h^{(1)}\,h^{(2)}+
  b_2\left(h^{(1)}\,h^{(2)}_{,ii}+h^{(2)}\,h^{(1)}_{,ii}\right)\nonumber\\&&+
  b_3\,h^{(1)}_{,i}\,h^{(2)}_{,i}+
  b_4\,h^{(1)}_{,ii}\,h^{(2)}_{,jj}+
  b_5\,h^{(1)}_{,ij}\,h^{(2)}_{,ij}\,.  
\end{eqnarray}
Expressing now $F$ in terms of $h$ and $u$, we obtain the decoupled
form $F=F_h+F_u$, with
\begin{eqnarray}
    F_h&=&
    d_1\,h^2+
    d_2\,h_{,i}\,h_{,i}+
    d_3\,h\,h_{,ii}+
    d_4\,h^2_{,ii}\nonumber\\&&+
    d_5\,h_{,ij}\,h_{,ij}\\
    F_u&=&
    e_1\, u+
    e_2\, u^2+
    e_3\,u_{,i}\, u_{,i}+
    e_4\, u_{,ii}+
    e_5\, u\, u_{,ii}\nonumber\\&&+
    e_6\,u^2_{,ii}+
    e_7\, u_{,ij}\, u_{,ij}\,.
\end{eqnarray}
The new coefficients are related to the former by an invertible linear
transformation. In this expression, several terms obviously vanish and
other can be discarded: $e_1\equiv0$, since the minimum energy
corresponds to $u=0$ by construction; $d_1=d_3\equiv0$ since $F$ must
be invariant under a translation. We shall set $d_2=0$, since the
tension of membranes usually vanishes~\cite{david,luca}.  There is no
reason however to discard the term $e_3(\nabla u)^2$, which represents
the energy density associated with a {\em gradient of the membrane
  thickness}. The latter term involves not only the extra coast of
lengthening the chain--water interfaces but also that of modulating
the stretching of the molecules chains. Considers a planar membrane
with a thickness modulation at some wavevector $q$: its elastic energy
may be well described by the term $\propto u^2$ as long as $qa\ll1$
($a$ monolayer thickness), however the term $\propto(\nabla u)^2$
should not be neglected when $qa\approx1$.  From this point of view
the present model differs from those of
Refs.~\cite{huang,dan1,dan2,dan3} that neglect the coefficient
$e_3$~\cite{note}.

We can now rewrite $F$ in a more traditional way. Relabeling the
nonzero coefficients, and making use of
$h_{,ij}\,h_{,ij}=(\nabla^2h)^2-2\,{\rm Det}(h_{,ij})$, we arrive at
  \begin{eqnarray}
    F_h&=&
    \frac{1}{2}\kappa\,(\nabla^2h)^2+
    \bar\kappa\,{\rm Det}(h_{,ij})\,,\\
    F_ u&=&
    \frac{1}{2}B\,u^2+
    \frac{1}{2}\lambda\left(\nabla u\right)^2+
    \sigma\,\nabla^2 u+
    \sigma'\,u\,\nabla^2 u\nonumber\\&&+
    \frac{1}{2}\kappa'\,(\nabla^2 u)^2+
    \bar\kappa'\,{\rm Det}(u_{,ij})\,.  
\end{eqnarray}
$F_h$ is simply the Helfrich energy~\cite{canhel1,canhel2}, in which
$\nabla^2h$ is twice the mean curvature of the average membrane shape,
and ${\rm Det}(h_{,ij})$ is its Gaussian curvature. The thickness
variations, which are completely decoupled from the average membrane
shape, are described by an energy $F_u$ similar to that of
Refs.~\cite{dan1,dan2,dan3}, however with two important differences: (1) there is a non
vanishing term $\propto\!\left(\nabla u\right)^2$ at
lowest-order, (2) the bending constants $\kappa'$ and $\bar\kappa'$
are different from the Helfrich constants appearing in $F_h$.

To further simplify our model, we shall discard the terms proportional
to $\sigma$ and $\sigma'$, since they can be transformed to boundary
terms by integration by parts; we shall also discard the terms
proportional to $\kappa'$ and $\bar\kappa'$, in order to keep only the
leading-order saturation terms. We are left (at the moment) with
\begin{equation}
F=\frac{1}{2}\kappa\,(\nabla^2h)^2+
    \bar\kappa\,{\rm Det}(h_{,ij})+
\frac{1}{2}B\,u^2+
    \frac{1}{2}\lambda\left(\nabla u\right)^2
\end{equation}

\subsection{Tilt distortion energy and coupling terms}

We expand the distortion energy associated with the tilts of the molecular
orientation in powers of
\begin{equation}
  m_i^{(\alpha)};\quad m_{i,j}^{(\alpha)};
\end{equation}
The tilt gradient $m^{(\alpha)}_{i,j}$ is a non-symmetric second rank
tensor. We write the tilt free energy as
$G=G^{(1)}+G^{(2)}+G^{(12)}+G_{\rm int}$ where all the terms coupling
${\bf m}^{(1)}$ and ${\bf m}^{(2)}$ are in $G^{(12)}$ and all the
interaction terms coupling the tilts and the membrane shape or
dilation are in $G_{\rm int}$. The interaction can be divided into
four contributions: $G_{\rm
  int}=G_u^{(1)}+G_u^{(2)}+G_h^{(1)}+G_h^{(2)}$, each term containing
all the contributions bilinear in either $m^{(1)}$ or $m^{(2)}$ and
$u$ or $h$. Because of the symmetry of the bilayer, we require
invariance with respect to the transformation (\ref{sym}) and the exchange
of ${\bf m}^{(1)}$ and ${\bf m}^{(2)}$. Writing all the linear and
quadratic scalars yields
  \begin{eqnarray}
    G^{(\alpha)}&=&
    A_1\,m_{i,i}^{(\alpha)}+
    A_2\,m_i^{(\alpha)}\,m_i^{(\alpha)}+
    A_3\,m_{i,i}^{(\alpha)2}\nonumber\\&&+
    A_4\,m_{i,j}^{(\alpha)}\,m_{i,j}^{(\alpha)}+
    A_5\,m_{i,j}^{(\alpha)}\,m_{j,i}^{(\alpha)}\\
    G^{(12)}&=&
    B_1\,m_i^{(1)}\,m_i^{(2)}+
    B_2\,m_{i,i}^{(1)}\,m_{j,j}^{(2)}\nonumber\\&&+
    B_3\,m_{i,j}^{(1)}\,m_{i,j}^{(2)}+B_4\,
      m_{i,j}^{(1)}\,m_{j,i}^{(2)}\\
  G_u^{(\alpha)}&=&
  C_1\,m_i^{(\alpha)}\,u_{,i}+
  C_2\,m_{i,i}^{(\alpha)}\,u+
  C_3\,m_{i,i}^{(\alpha)}\,u_{,jj}\nonumber\\&&+
  C_4\,m_{i,j}^{(\alpha)}\,u_{,ij}\\
  G_h^{(\alpha)}&=&
  (-1)^{\alpha}D_1\,m_i^{(\alpha)}\,h_{,i}+
  (-1)^{\alpha}D_2\,m_{i,i}^{(\alpha)}\,h\nonumber\\&&+
  (-1)^{\alpha}D_3\,m_{i,i}^{(\alpha)}\,h_{,jj}+
  (-1)^{\alpha}D_4\,m_{i,j}^{(\alpha)}\,h_{,ij}
\end{eqnarray}
As previously, several terms can be discarded: $A_1=C_2=D_2\equiv0$,
since we assume no spontaneous splay of the tilt; $D_1\equiv0$ since
the minimum energy is still achieved with zero tilts when the membrane
is rotated. We can also discard the terms with coefficients $A_5$ and
$B_4$: integrating them by parts merely yields boundary terms (and a
renormalization of $A_3$ and $B_2$).

In terms of the variables ${\bf m}$ and ${\bf\widehat m}$, we can write
the total tilt energy as $G=G_{\bf m}+G_{{\bf m}h}+G_{\bf\widehat
  m}+G_{{\bf\widehat m}u}$, with
\begin{eqnarray} 
    G_{\bf m}&=&
    \frac{1}{2}t\,m_i\,m_i+
    k_1\,m_{i,i}^2+
    k_2\,m_{i,j}\,m_{i,j}\\
    G_{{\bf m}h}&=&
    d_3\,m_{i,i}\,h_{,jj}+
    d_4\,m_{i,j}\,h_{,ij}\\
    G_{\bf\widehat m}&=&
    \frac{1}{2}t'\,\widehat m_i\,\widehat m_i+
    k'_1\,\widehat m_{i,i}^2+
    k'_2\,\widehat m_{i,j}\,\widehat m_{i,j}\\
    G_{{\bf\widehat m}u}&=&
    c\,\widehat m_i\,u_{,i}+
    c_1\,\widehat m_{i,i}\,u_{,jj}+
    c_2\,\widehat m_{i,j}\,u_{,ij}\,. 
\end{eqnarray}
We can finally discard the terms with coefficients $c_2$ and $d_4$ by
integrating by parts, and neglect the term with coefficient $c_1$ as a
higher-order coupling term.

\subsection{Total distortion energy}

\begin{figure}
  \resizebox{0.5\textwidth}{!}
  {\hspace{100pt}\includegraphics{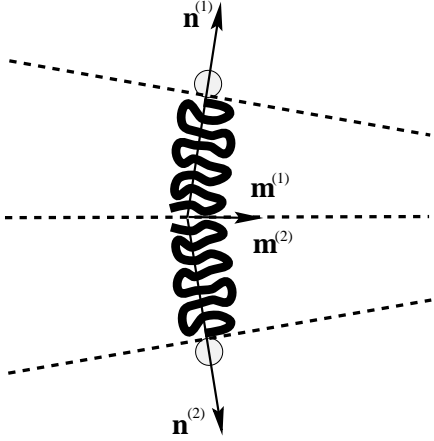}\hspace{100pt}}
  \caption{Coupling between the tilt-difference 
    ${\bf\widehat m}=\frac{1}{2}({\bf m}_1+{\bf m}_2)$ and the
    thickness gradient $\nabla u$, via the term $c\,\nabla
    u\cdot{\bf\widehat m}$.}
  \label{fig:coupling}
\end{figure}

In vectorial notations and after some simple manipulations, the total
distortion energy, i.e., $F+G$, can be written as $H_{hm}+H_{u\widehat
  m}$, with
\begin{eqnarray}
  H_{hm}&=&
  \frac{1}{2}\kappa\,(\nabla^2h)^2+
  \bar\kappa\,{\rm Det}\,(h_{,ij})
  -\gamma\,\nabla^2 h\,(\nabla\cdot{\bf m})\nonumber\\&+&
  \frac{1}{2}t\,{\bf m}^2+
  \frac{1}{2}K_1\,(\nabla\cdot{\bf m})^2+
  \frac{1}{2}K_2\,(\nabla\times{\bf m})^2
\end{eqnarray}
and
\begin{eqnarray}\label{umc}
  H_{u\widehat m}&=&
  \frac{1}{2}B\,u^2+
  \frac{1}{2}\lambda\left(\nabla u\right)^2+
  c\,\nabla u\cdot{\bf\widehat m}\nonumber\\&+&
  \frac{1}{2}t'\,{\bf\widehat m}^2+
  \frac{1}{2}K'_1\,(\nabla\cdot{\bf\widehat m})^2+
  \frac{1}{2}K'_2\,(\nabla\times{\bf\widehat m})^2
\end{eqnarray}
The total energy therefore splits up into a contribution $H_{hm}$
involving the average shape $h$ and the average tilt ${\bf m}$, and a
decoupled contribution $H_{u\widehat m}$ involving the dilation $u$
and the tilt-difference ${\bf\widehat m}$.  The term with coefficient
$\gamma>0$ is responsible for the ripple phase of tilted
membranes~\cite{ripple1,ripple2}. Similarly, the term with coefficient
$c$ can produce a ``ripple'' instability in which a thickness
modulation occurs together with a tilt-difference
modulation~\cite{euro}. From the tendency of the molecules to orient
perpendicular to the chain--water interface we expect $c>0$
(Fig.~\ref{fig:coupling}).

\subsection{Equilibrium equations and energies}

The total elastic energy of the membrane is given by
\begin{equation}\label{totalnrj}
  {\cal H}={\cal H}_{hm}+{\cal H}_{u\widehat m}=
  \int\!{\rm d}^2r\,H_{hm}+
  \int\!{\rm d}^2r\,H_{u\widehat m}.
\end{equation}
The equilibrium membrane configuration are those minimizing ${\cal H}$
with respect to all possible local variations of the structural
fields. The four corresponding Euler-Lagrange equations, namely
$\delta{\cal H}/\delta h=0$, $\delta{\cal H}/\delta{\bf
  m}=0$, $\delta{\cal H}/\delta u=0$ and
$\delta{\cal H}/\delta {\bf\widehat m}=0$, are explicitly
\begin{eqnarray}
  \label{hta}
  \kappa\,\nabla^4h&=&
  \gamma\,\nabla^2\left(\nabla\cdot{\rm m}\right)\\
  \label{htb}
  t\,{\bf m}-K_1\,\nabla\left(\nabla\cdot{\bf m}\right)&+&
  K_2\,\nabla\!\times\!\left(\nabla\!\times\!{\bf m}\right)
  =-\gamma\,\nabla\left(\nabla^2h\right)\nonumber\\
\end{eqnarray}
and
\begin{eqnarray}
  \label{uaa}
  &B\,u-\lambda\,\nabla^2u=
  c\,\nabla\cdot{\bf\widehat m}&\\
  \label{uab}
  &t'\,{\bf\widehat m}-K'_1\,\nabla\left(\nabla\cdot{\bf\widehat m}\right)+
  K'_2\,\nabla\!\times\!\left(\nabla\!\times\!{\bf\widehat m}\right)=
  -c\,\nabla u&\!\!.
\end{eqnarray}
The calculation of the energy of equilibrium configurations can be
simplified in the following way. Integrating ${\cal H}_{u{\widehat
    m}}$ by parts yields
\begin{equation}\label{stokes}
{\cal H}_{u{\widehat m}}=\frac{1}{2}\int\!{\rm d}^2r\,
\left(u\,\frac{\delta{\cal H}}{\delta u}+
{\bf\widehat m}\cdot\frac{\delta{\cal H}}{\delta{\bf\widehat m}}\right)
\,\,+\,\,{\cal H}'_{u{\widehat m}}
\end{equation}
with
\begin{eqnarray}\label{eq:energie.up}
 {\cal H}'_{u{\widehat m}}=
 \frac{1}{2}\!\oint\!d\ell\,{\bf n}&\cdot&\left[
    \lambda\,u\nabla u+
    c\,u\,{\bf\widehat m}+
    K'_1\left(\nabla\cdot{\bf\widehat m}\right)\,{\bf\widehat m}
  \right.\nonumber\\&&-\left.
    K'_2\left(\nabla\!\times\!{\bf\widehat m}\right)\!
\times\!{\bf\widehat m}
  \right]\,,
\end{eqnarray}
where the last integral is restricted to the boundary of the
integration domain, whose normal is ${\bf n}$. For equilibrium
configurations, ${\cal H}_{u{\widehat m}}$ reduces to ${\cal
  H}'_{u{\widehat m}}$ since the first term of~(\ref{stokes})
vanishes. This provides a very useful simplification.
  
  On finds similarly that ${\cal H}_{hm}$ reduces for equilibrium
  configurations to
\begin{eqnarray}
  {\cal H}'_{hm}&=&\frac{1}{2}\!\oint\!d\ell\,{\bf n}\cdot\left[
    \kappa\,\nabla^2h\,\nabla h-
    \kappa\,h\,\nabla\!\left(\nabla^2h\right)\right.\nonumber \\&-&\left.
    \gamma\,\left(\nabla\cdot{\bf m}\right)\nabla h+
    \gamma\,h\,\nabla\!\left(\nabla\cdot{\bf m}\right)-
    \gamma\,\left(\nabla^2h\right){\bf m}\right.\nonumber\\&+&\left.
    K_1\left(\nabla\cdot{\bf m}\right)\,{\bf m}-
    K_2\left(\nabla\times{\bf m}\right)\times{\bf m}
  \right]\nonumber\\\label{calH}
\end{eqnarray}

\subsection{Orders of magnitude}\label{sec:oom}

For biological membranes, the bending constants $\kappa>0$ and
$\bar\kappa<0$ have relatively high values $\simeq10^{-12}\,{\rm erg}$
($\simeq25\,$ $k_{\rm B}T$) \cite{isra}. The typical value of the
membrane area-stretching coefficient $k\simeq100\,{\rm erg}/{\rm
  cm}^2$~\cite{isra} allows to determine the dilation modulus via
$B=k/(2a)^2$, where $a\simeq20\times10^{-8}\,{\rm cm}$ is a typical
monolayer thickness. This yields $B\simeq6\times10^{14}\,{\rm
  erg}/{\rm cm}^4$. Therefore $B\simeq\kappa/a^4$: the membrane
has a typical energy scale given by $\kappa$ and a typical length
scale given by $a$. In the absence of experimental measurements, the
other constants have to be estimated by dimensional analysis. We
expect $\lambda$ to be $\approx\kappa/a^2$ (we recall that $\lambda$
is independent of the membrane tension). We therefore estimate
$\lambda\approx25\,{\rm erg}/{\rm cm}^2$ Next, we shall assume roughly
that tilting the molecules by a large angle compares energetically
with compressing the membrane by half a monolayer thickness. This
yields $t\approx t'\approx\lambda$. Then, we expect the characteristic
length defined by the $(K_i/t)^{1/2}$ to be of order $a$, which
implies $K_i\approx\lambda a^2\approx10^{-12}\,{\rm erg}$. This value
correctly compares with the bending constant.  Finally, the $K'_i$'s
are expected to be of the same order of magnitude as the $K_i$'s, and
$c$ is dimensionally expected to compare with $\lambda$.

\subsection{Remarks on the validity of the truncation
  of the energy expansion}\label{sec:trunc}

Strictly speaking, in all the microscopic theories of
membranes~\cite{owicki,huang,dan1,dan2,dan3}, it is somewhat arbitrary to
truncate the expansion at the lowest-order in the derivatives of the
distortion field. Indeed, since the typical energy and
length scales of the membrane are $\kappa$ and $a$, respectively, the
distance $\xi$ on which dilation perturbations relax is expected to
be $\approx a$.  For small distortions $u\!\ll\!a$ there is no
problem in neglecting quartic terms such as the one $\sim\!(\nabla
u)^4$: this term is $(u/a)^2$ times smaller than the leading term
$\sim\!(\nabla u)^2$.

However the term $\sim\!(\nabla^2u)^2$, which we have discarded, might
be of the same order of magnitude as the leading term $\sim\!(\nabla
u)^2$ if indeed its coefficient is $\simeq\kappa$.  The problem is
that all the terms $\sim\!(\nabla^{n}u)^2$ may be also comparable if
their coefficients are $\simeq\kappa\,a^{2n-4}$.  However, at the
microscopic scale corresponding to $a$, the membrane is not actually a
continuum and there is not much meaning in considering high order
derivatives of the thickness. It may therefore be a good approximation
to keep only the leading order term.

In any case, we expect that the lowest-order truncation of this
continuum description will give a correct physical picture, at least
qualitatively, of the competing trends associated with the various
elastic variables. Note also that the truncation may be technically
correct in the vicinity of a transition to a more ordered $L_\beta$ or
$L_\beta'$ phase with a different equilibrium thickness, where $B$
might be significantly reduced and, accordingly, $\xi$ larger than
$a$.

\section{Interactions among membrane inclusions}

Biological membranes contain a large number of inclusions such as
integral proteins. Inclusions with a conical shape tend to curve the
membrane since the lipids orient parallel to the inclusion's boundary
in order to fill the volume. Because of the interference between the
resulting membrane distortions, such inclusions are subject to
long-range interactions~\cite{goulian,park,netz}.  Inclusions also
experience ``Casimir'' forces, which are due to the modification of
the membrane fluctuation spectrum caused by their presence.
 
The {\em short-range} interactions between inclusions arise from the
local structural changes that the latter impose on the
membrane~\cite{marcel,owicki,huang,dan1,dan2,dan3}. For instance, since proteins
have a central hydrophobic region that spans the hydrophobic core of
the membrane, a thickness mismatch between the hydrophobic region of
the protein and that of the bilayer will result in a local membrane
thickness perturbation. Interferences between such perturbations yield
membrane-mediated interactions that add up to the standard
screened-electrostatic and van der Waals interactions.

\subsection{Boundary conditions}

\begin{figure}
  \resizebox{0.5\textwidth}{!}
  {\hspace{5pt}\includegraphics{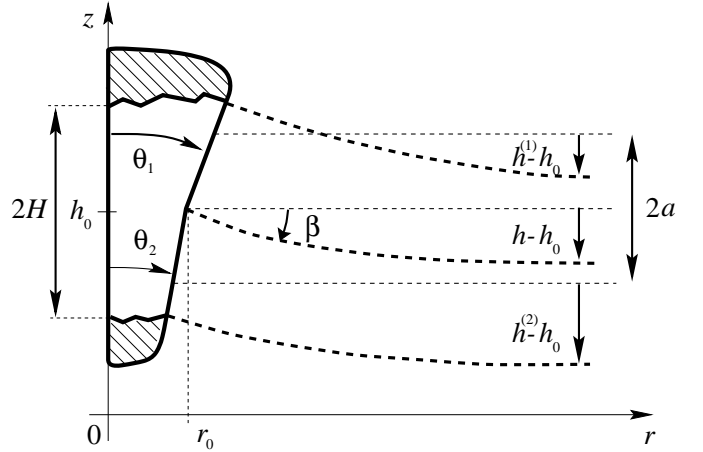}\hspace{5pt}}
\caption{
  General boundary conditions imposed by an inclusion.}
\label{fig:general}
\end{figure}

Let us consider a membrane inclusion such as the one depicted in
Fig.~\ref{fig:general}.  It is meant to model an integral protein with
an arbitrary shape.  For the sake of simplicity, however, we assume
revolution symmetry. We suppose that the hydrophobic region of the
inclusion has a thickness $2H$ that differs from the corresponding
thickness $2a$ in the bilayer. The inclusion is also assumed to have a
piecewise conical shape with two angles $\theta_1$ and $\theta_2$
pertaining to each monolayer and relative to the revolution axis.

Let us consider an undistorted reference membrane above which the
inclusion stands at a height $h_0$. As previously we denote by
$h^{(1)}$ and $h^{(2)}$ the positions of the upper and lower membrane
interfaces with respect to their equilibrium positions in the
reference membrane. Assuming a strong coupling between hydrophobic
parts~\cite{owicki,huang,dan1,dan2,dan3}, we require the conditions that both
monolayers interfaces reach the inclusion at the separation line
between its hydrophobic and hydrophilic regions, i.e.,
\begin{eqnarray}
  h^{(1)}|_{r_0}&\simeq&h_0+H-a\,,\\
  h^{(2)}|_{r_0}&\simeq&h_0-(H-a)\,.
\end{eqnarray}
These conditions are only approximate because the position where the
interfaces reach the inclusion is equal to $r_0$ only at lowest order
in the deformation variables.  Another boundary ``condition'', which
is not imposed but actually free to adjust to equilibrium, is the
angle $\beta$ at which the mid-membrane shape $h$ departs from the
inclusion. Calling ${\bf e}_{r}$ the unit vector along $r$, this
condition is
\begin{equation}
  \nabla h|_{r_0}\simeq\beta\,{\bf
  e}_{r}\,.
\end{equation}
If we now require that the molecules within the membrane lie parallel
to the inclusion's boundary, because of the space-filling constraint, we
have the condition
\begin{eqnarray}
  {\bf p}^{(1)}|_{r_0}&\simeq&-\theta_1\,{\bf
  e}_{r}\,,\\
  {\bf p}^{(2)}|_{r_0}&\simeq&\theta_2\,{\bf
  e}_{r}\,.
\end{eqnarray}
Note that we have implicitly assumed that the revolution axis of the
inclusion is normal to the reference plane $(x,y)$, although in the
most general situation it can be a tilted (this tilt will be zero by
symmetry in the following).

\subsubsection{Decoupled boundary conditions}\label{sec:decoupled}

In order to make use of the equilibrium equations previously derived,
we must transform these boundary conditions into conditions involving
the variables $h$, $u$, ${\bf m}$, and ${\bf\widehat m}$. From
Eqs.~(\ref{def_h}--\ref{def_u}) and
Eqs.~(\ref{def_p}--\ref{def_pchap}), we obtain
\begin{eqnarray}
  \label{bch}
  h|_{r_0}&\simeq& h_0\,,\\
  \nabla h|_{r_0}&\simeq& \beta\,{\bf
  e}_{r}\,,\\
  \label{bct}
  \vec{m}|_{r_0}&\simeq& \left(\beta-\Theta\right){\bf
  e}_{r}\,,
\end{eqnarray}
and
\begin{eqnarray}
  \label{bcu}
  u|_{r_0}&\simeq& u_0\,,\\
  \label{bca}
  \vec{\widehat m}|_{r_0}&\simeq&\alpha_0\,{\bf
  e}_{r}\,,
\end{eqnarray}
where $\Theta=\frac{1}{2}(\theta_1+\theta_2)$ is the average cone
angle of the inclusion, $u_0=H-a$ is the dilation and
$\alpha_0=\frac{1}{2}(\theta_2-\theta_1)$ the tilt-difference set by
the inclusion. It is important to note that these two sets of boundary
conditions are decoupled in the same way as the corresponding
equilibrium equations.

\subsubsection{Arrays of inclusions}

\begin{figure}
  \resizebox{0.5\textwidth}{!}
  {\hspace{100pt}\includegraphics{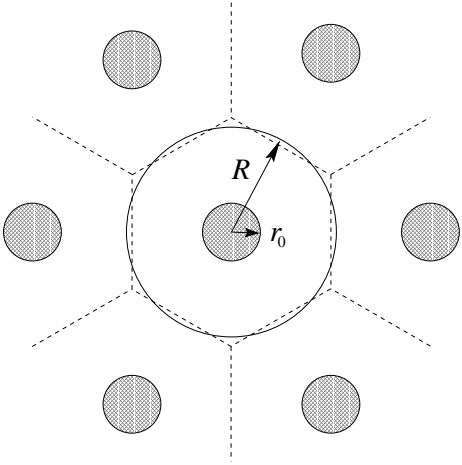}\hspace{100pt}}
\caption{An array of inclusions and its Wigner-Seitz cell.}
\label{fig:wigner}
\end{figure}

Following previous works, we shall calculate the constitutive energy
of {\em arrays} of inclusions. Paradoxally, it is easier to
approximatively calculate the energy of an array than to calculate
exactly the interaction between two inclusions. Since
membrane-mediated interactions are not pairwise additive, this is the
correct procedure to investigate the stability of $2D$ crystalline
structures.

To capture the physics of an array of inclusions, the standard method
is to consider a single inclusion surrounded by its Wigner-Seitz cell
(i.e., the unit cell made by the perpendicular bisectors of the
bonds connecting the lattice sites)~\cite{owicki,dan1,dan2,dan3}.  The
Wigner-Seitz cell is further idealized by a circle of radius
approximatively half the inclusions separation (cf.\ 
Fig.~\ref{fig:wigner}), and the equilibrium equations are solved
assuming {\em revolution symmetry}, with boundary conditions at
$r=r_0$ and $r=R$. When applied to an hexagonal lattice of inclusions,
this approximation is quite good, as it consists in neglecting high
Fourier harmonics of order $6$, $12$, etc. In a gas of inclusions,
it amounts to considering that the first neighbors effectively
screen the other inclusions.

\subsection{Dilation--tilt-difference induced interactions
  in an array of inclusions}

Inclusions with arbitrary shapes will in general excite all of the
four distortion modes considered in this work. However, since we have
seen that both the equilibrium equations and the boundary conditions
are pairwise decoupled, one can study separately, and simply add, the
effects of the coupled dilation and tilt-difference modes and the
effects of the coupled shape and tilt modes.

\subsubsection{Zero dilation--tilt-difference coupling}

We focus on the dilation ($u$) and tilt-difference (${\bf\widehat m}$)
modes and, to start with, we neglect their coupling:
\begin{equation}
  c=0\,.
\end{equation}
Let us consider an array of inclusions, and assume, as previously
discussed, a perfect revolution symmetry in the Wigner-Seitz cell
surrounding an inclusion:
\begin{equation}\label{eq:revup}
  u=u(r) \quad{\rm and}\quad \vec{\widehat m}=\alpha(r)\,{\bf
  e}_{r}\,.
\end{equation}
Under these conditions, the most general solution of the equilibrium
equations~(\ref{uaa}-\ref{uab}) takes the form
\begin{eqnarray}
  \label{2xiu}
  u(r)&=&\left[
    A_1\,\,{\rm K}_0\!\left({r\over\xi_u}\right)+
    A_2\,\,{\rm I}_0\!\left({r\over\xi_u}\right)
  \right]\times\sqrt{t'\over B}\,,\\
  \label{2xia}
  \alpha(r)&=&\left[
    A_3\,\,{\rm K}_1\!\left({r\over\xi_\alpha}\right)+
    A_4\,\,{\rm I}_1\!\left({r\over\xi_\alpha}\right)
  \right]\,,
\end{eqnarray}
in which the I's and the K's are modified Bessel function and
\begin{eqnarray}
\xi_u&=&\sqrt{\frac{\lambda}{B}}\,,\\
\xi_\alpha&=&\sqrt{\frac{K'_1}{t'}}\,,
\end{eqnarray}
are two characteristic length comparable with the membrane thickness,
except close to a $L_\beta$ tilted phase where $t'$ might be small, or
close to the main-chain transition where $B$ might be small. The
constants $A_{\rm i}$'s, which are real and dimensionless, are
determined from the boundary conditions:
\begin{eqnarray}
\label{bd}
  u|_{r_0}&=&u_0\,,\\
  \alpha|_{r_0}&=&\alpha_0\,,\\
  \left.\dot u\right|_R&=&0\,,\\
\label{bf}
  \left.\alpha\right|_R&=&0\,,
\end{eqnarray}
with a dot indicating derivation with respect to $r$. The quantities
$u_0$ and $\alpha_0$ are the boundary dilation and tilt-difference,
respectively. The last two conditions are required by symmetry on the
Wigner-Seitz circle.

\begin{figure}
  \resizebox{0.5\textwidth}{!}
  {\hspace{40pt}\includegraphics{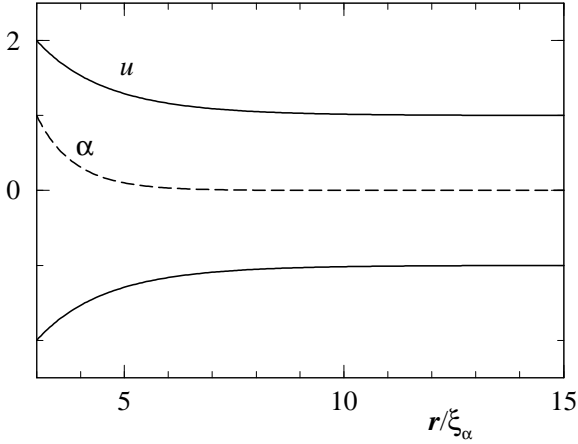}\hspace{40pt}}
\caption{
  Sketch of the membrane structure around an isolated inclusion
  (distortions are amplified, see text). The inclusion radius is
  $r_0=3\,\xi_\alpha(\simeq60\,{\rm\AA})$, $\xi_u/\xi_\alpha=2$ and
  $c=0$.}
\label{fig:mem.c0}
\end{figure}

\begin{figure}
  \resizebox{0.5\textwidth}{!}
  {\hspace{40pt}\includegraphics{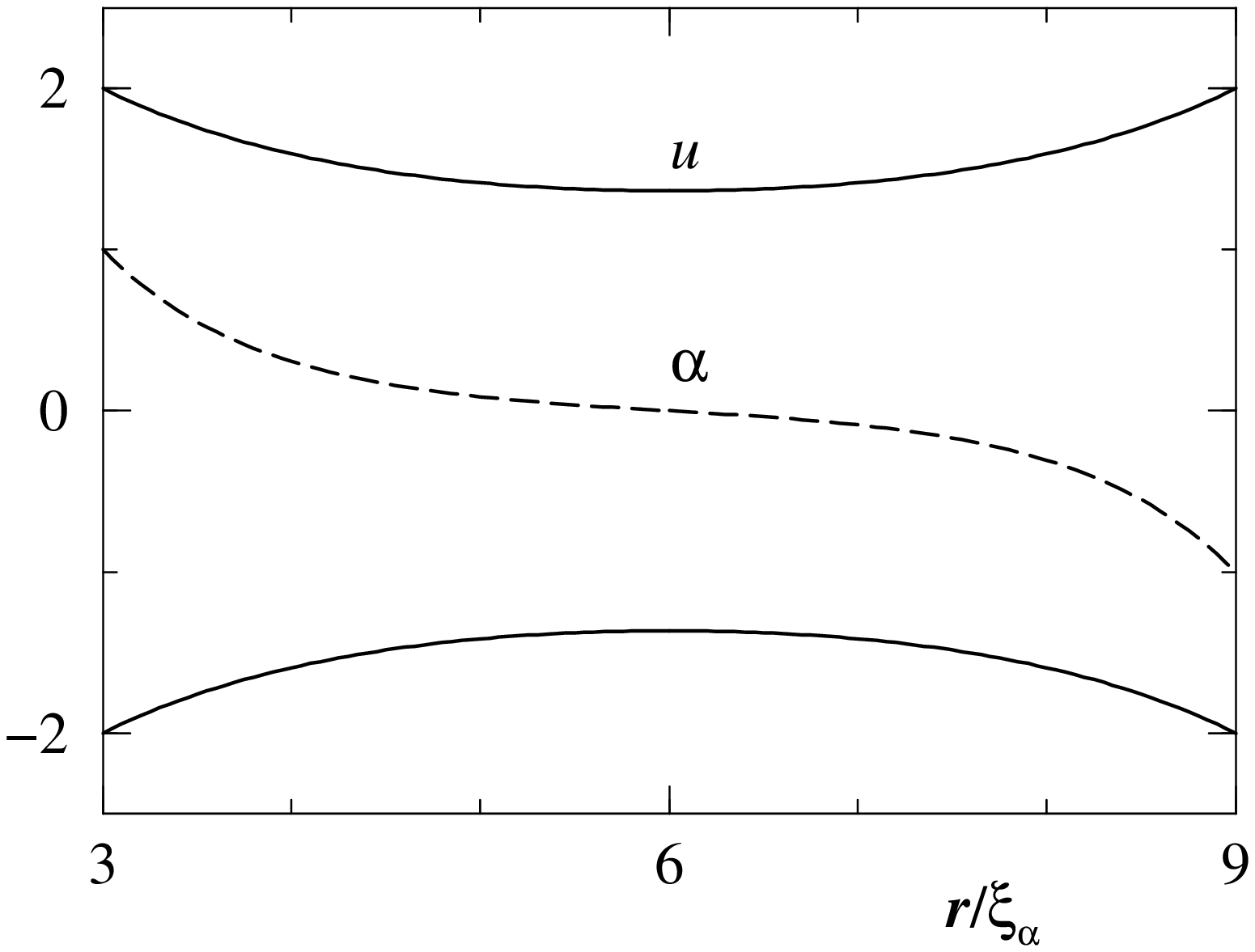}\hspace{40pt}}
\caption{
  Membrane structure between two inclusions in the
  array (amplified distortions). Parameters are as in
  Fig.~\ref{fig:mem.c0}.}
\label{fig:proch0}
\end{figure}

Figure~\ref{fig:mem.c0} shows a typical solution for an isolated
inclusion ($R\!\to\!\infty$) and Fig.~\ref{fig:proch0} shows a typical
solution corresponding to an array of interacting inclusions. These
pictures sketch the membrane structure: the solid line represents the
membrane shape, i.e., the sum of the equilibrium monolayer thickness
$a$ and the thickness excess $u$. The dashed curve represents the
amplitude of the tilt-difference angle $\alpha$. For the sake of
clarity the distortions have been amplified in the following way: the
boundary angle $\alpha_0$, the equilibrium monolayer thickness $a$, and
the boundary thickness excess $u_0$ are all normalized to $1$.


\begin{figure}
  \resizebox{0.5\textwidth}{!}
  {\hspace{40pt}\includegraphics{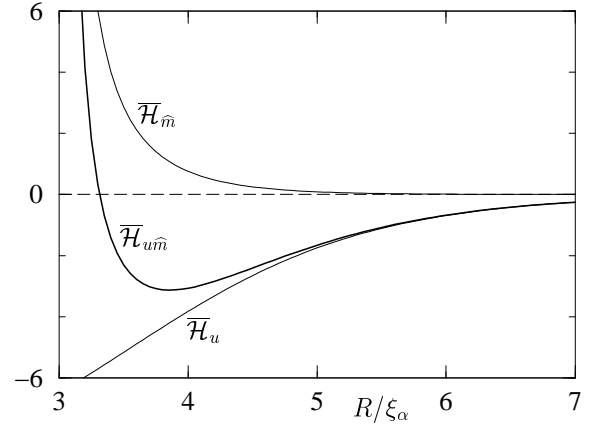}\hspace{40pt}}
\caption{
  Normalized interaction energy per inclusion $\overline{\cal
    H}_{u\widehat m}$ {\em vs.}~the inclusions separation $R$. The
  curves correspond to $r_0=3\,\xi_\alpha(\simeq60\,{\rm\AA})$, $s=2$,
  $x=1$ and $c=0$. The normalized energies $\overline{\cal H}_{u}$ and
  $\overline{\cal H}_{\widehat m}$ correspond to the attractive
  dilation and repulsive tilt-difference contributions, respectively.}
\label{fig:nrj.c0}
\end{figure}

Assuming revolution symmetry, the general distortion
energy~(\ref{eq:energie.up}) within the Wigner-Seitz cell, takes the
form
\begin{equation}\label{eq:energie.up.radial}
{\cal H}_{u\widehat m}=
\pi\left[\lambda\,r\,u\,\dot u+c\,r\,u\,\alpha+K'_1\,r\,\alpha\,\dot\alpha
+K'_1\,\alpha^2\right]_{r_0}^R\,,
\end{equation}
in which several terms vanish due to the boundary conditions. After
eliminating constant terms, ${\cal H}_{u\widehat m}$ reduces to
\begin{equation}\label{eq:energie.up2}
{\cal H}_{u\widehat m}=
{\cal H}_{u}+{\cal H}_{\widehat m}=
-\pi r_0\left(\lambda\,u_0\,\dot u|_{r_0}
+K'_1\,\alpha_0\,\dot\alpha|_{r_0}\right)\,.
\end{equation}
This interaction, which in principle depends on a large number of
parameters ($r_0$, $R$, $u_0$, $\alpha_0$, $B$, $\lambda$, $t'$ and
$K'$) has the following scaling property:
\begin{eqnarray}
\frac{{\cal H}_{u\widehat m}}{\pi B\,r_0\,\xi_\alpha u_0^2}
&=&\overline{\cal H}_{u\widehat m}\left(
x^2,s,\frac{r_0}{\xi_\alpha},\frac{R}{\xi_\alpha}
\right)\,,\label{eq:nor}\\
s&=&\frac{\xi_u}{\xi_\alpha}\,,\\
x&=&\frac{\alpha_0}{u_0\sqrt{B/t'}}\,,
\end{eqnarray}
which advantageously reduces the effective numbers of parameters.  At
short inclusions separations, ${\cal H}_{\widehat m}$ diverges as
$(R-r_0)^{-1}$ and ${\cal H}_{u}$ goes to a negative constant. At large
separations, both relax exponentially. Figure~\ref{fig:nrj.c0} shows a
typical situation where an energy minimum appears, resulting from the
superposition of a dilation-induced attraction, that dominates at
large distances, and a tilt-difference--induced repulsion, that
dominates at short distances. This situation manifests itself for
large values of $\xi_u$, for which the dilation mode has the longest
range, and for small values of the boundary tilt-difference
$\alpha_0$, for which the repulsion is weak

Let us estimate the magnitude of the interaction energy, which is
given by the normalization factor $\pi B\,r_0\,\xi_\alpha u_0^2$
in~(\ref{eq:nor}). We choose for the tilt-difference coherence length a fixed
microscopic value $\xi_\alpha\simeq20\,{\rm\AA}$ and we let for instance
$0.1<s<10$. This assumption is based on
the fact that close to the main chain transition $\xi_u$ should
exhibit some degree of pretransitional divergence. For the inclusion,
we assume a typical protein size $r_0=3\,\xi_\alpha(\simeq60\,{\rm\AA})$
and a thickness perturbation $u_0=0.2\,\xi_\alpha(\simeq4\,{\rm\AA})$.
With the estimated values of the material constants given in
Sec.~\ref{sec:oom}, we obtain $\pi B\,r_0\,\xi_\alpha
u_0^2\simeq(10/s^2)k_{\rm B}T$.

In the energy graphs depicted in Fig.~\ref{fig:nrj.c0}, the values
$x\!=\!1$ and $s\!=\!2$ correspond to an inclusion boundary
tilt-difference angle $\alpha_0=(x/s)(u_0/\xi_\alpha)
\sqrt{\lambda/t'}\simeq6^\circ$. The depth of the energy minimum is
$3\times\pi B\,r_0\,\xi_\alpha u_0^2\simeq7\,k_{\rm B}T$. For such a
well, we expect that the array of inclusion will crystallize, the
distance between the boundaries of the particles being then
$2(R-r_0)\simeq2\times0.8\,\xi_\alpha\simeq35\,{\rm\AA}$~\cite{par_surface}.

\begin{figure}
  \resizebox{0.5\textwidth}{!}
  {\hspace{40pt}\includegraphics{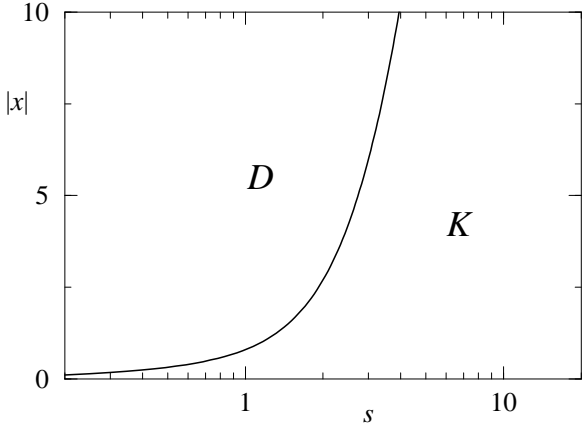}\hspace{40pt}}
\caption{
  Phase diagram for a membrane with $\xi_\alpha\simeq20\,{\rm\AA}$
  containing dilation-tilt-difference inducing inclusions with radius
  $r_0=3\,\xi_\alpha(\simeq60\,{\rm\AA})$. The coupling $c$ is
  neglected. (D) disordered phase. (K) crystal phase, as determined
  from the existence of an energy minimum deeper that $k_{\rm B}T$.}
\label{fig:diaph.c0}
\end{figure}

If we consider the inclusions radius $r_0$ as fixed, the interaction
potential as a function of $R$ depends only on the parameters $x$ and
$s$, as can be seen in~(\ref{eq:nor}). We have plotted in
Fig.~\ref{fig:diaph.c0} the phase diagram, in the $(x,s)$ plane, for
a collection of identical inclusions.  Distinction is made between a
disordered (D) gaseous state and a crystal (K) phase. The criterion
for the latter is the existence of a energy minimum with a depth
larger than $k_{\rm B}T$.

\subsubsection{Nonzero dilation--tilt-difference coupling}

In order to study the effect of the dilation--tilt-difference coupling, we
now assume
\begin{equation}
  c\ne0\,,\quad{\rm and}\quad\xi_u=\xi_\alpha\equiv\xi\,.
\end{equation}
The latter condition is a simplification, which is reasonable far from
any membrane phase transition (with $\xi$ of the order of the
membrane thickness). Let us define the coupling's characteristic length as
\begin{equation}
\ell=\frac{c}{2\sqrt{Bt'}}\,.
\end{equation}
We assume $\ell<\xi$, otherwise the membrane undergoes the microscopic
``ripple'' instability already mentioned~\cite{euro}.

\begin{figure}
  \resizebox{0.5\textwidth}{!}
  {\hspace{40pt}\includegraphics{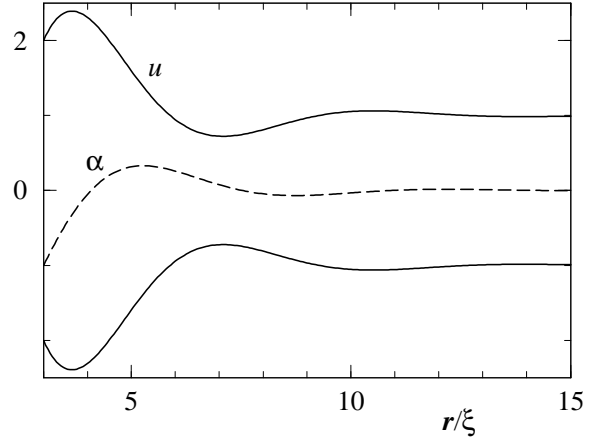}\hspace{40pt}}
\caption{
  Sketch of the membrane structure around an isolated inclusion
  (amplified distortions). The inclusion radius is
  $r_0=3\,\xi\,(\simeq60\,{\rm\AA})$, $x\!=\!-2$, and
  $\phi\!=\!0.75\times\pi/2$.}
\label{fig:mem.cne0}
\end{figure}

\begin{figure}
  \resizebox{0.5\textwidth}{!}
  {\hspace{40pt}\includegraphics{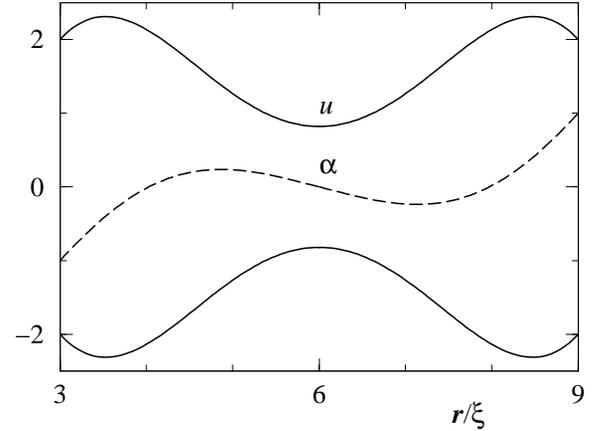}\hspace{40pt}}
\caption{
  Membrane structure between two inclusions in the
  lattice (amplified distortions). Parameters are as in
  Fig.~\ref{fig:mem.cne0}.}
\label{fig:proch1}
\end{figure}

Under the revolution symmetry conditions~(\ref{eq:revup}), the most
general solution of the equilibrium equations~(\ref{uaa}-\ref{uab})
is given by the real part of
\begin{eqnarray}
  \label{1xiu}
  u(r)&=&\left[
    \tens{A}_1\,\,{\rm K}_0\!\left({\rm e}^{{\rm i}\phi}\,{r\over\xi}\right)+
    \tens{A}_2\,\,{\rm I}_0\!\left({\rm e}^{{\rm i}\phi}\,{r\over\xi}\right)
  \right]\times\sqrt{t'\over B}\,,\\
  \label{1xia}
  \alpha(r)&=&\left[
    \tens{A}_1\,\,{\rm K}_1\!\left({\rm e}^{{\rm i}\phi}\,{r\over\xi}\right)-
    \tens{A}_2\,\,{\rm I}_1\!\left({\rm e}^{{\rm i}\phi}\,{r\over\xi}\right)
  \right]\times{\rm i}\,,
\end{eqnarray}
where ${\rm i}=\sqrt{-1}$, $\tens{A}_1$ and $\tens{A}_2$ are two
dimensionless {\em complex} constants, and
\begin{equation}
  \sin\phi={\ell\over\xi}\,.
\end{equation}
The constants $\tens{A}_1$ and $\tens{A}_2$ are determined from the
boundary conditions~(\ref{bd}--\ref{bf}) as previously.
Figures~\ref{fig:mem.cne0} and~\ref{fig:proch1} show a typical
solution for an isolated inclusion ($R\!\to\!\infty$) and a typical
solution for interacting inclusions, respectively. The same
conventions as for Figs.~\ref{fig:mem.c0} and~\ref{fig:proch0} are
used.

\begin{figure}
  \resizebox{0.5\textwidth}{!}
  {\hspace{40pt}\includegraphics{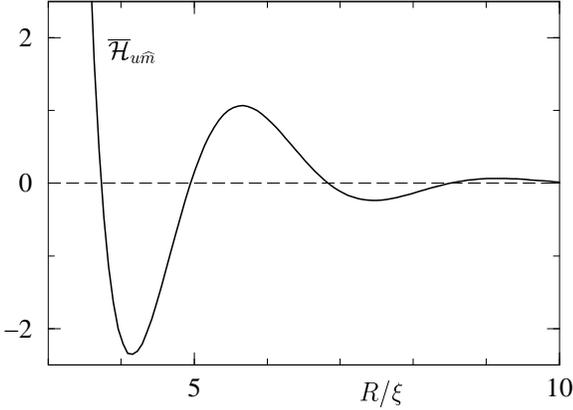}\hspace{40pt}}
\caption{
  Normalized interaction energy per inclusion $\overline{\cal
    H}_{u\widehat m}$ {\em vs.}~the inclusions separation $R$. The
  curves correspond to $r_0=3\,\xi\,(\simeq60\,{\rm\AA})$, $x\!=\!-3$,
  and $\phi\!=\!0.75\times\pi/2$.}
\label{fig:nrj.cne0}
\end{figure}

The distortion energy within the Wigner-Seitz cell is given by exactly
the same formula~(\ref{eq:energie.up2}) as previously. Indeed,
although the second term of~(\ref{eq:energie.up.radial}) does not
vanish any longer in $r\!=\!r_0$, it is constant and can be omitted in
the interaction. Note however that this energy cannot be splitted any
longer into pure dilation and tilt-difference contributions. Again,
${\cal H}_{u\widehat m}$ has the following scaling property:
\begin{equation}
  \frac{{\cal H}_{u\widehat m}}{\pi B\,r_0\,\xi\,u_0^2}=
\overline{\cal H}_{u\widehat m}\left(
    x,\phi,\frac{r_0}{\xi},\frac{R}{\xi}
  \right)\,.\label{eq:nor2}
\end{equation}
Depending on the values of $r_0/\xi$, $x$ and $\phi$, the interaction
energy is either monotonically repulsive or exhibits one {\em or
  several} marked minima. Figure~\ref{fig:nrj.cne0} shows a typical
situation in which two minima appear. This phenomenon manifests itself
for values of the dilation--tilt-difference coupling corresponding to
$\phi\!>\!0.6\times\pi/2$, where, because of the vicinity of the
dilation--tilt-difference ``ripple'' instability, the
membrane has a tendency to develop damped undulations.

The magnitude of the interaction energy is now given by the
normalization factor $\pi B\,r_0\,\xi\,u_0^2$.  With typically
$\xi\simeq20\,{\rm\AA}$, and again $r_0=3\,\xi\,(\simeq60\,{\rm\AA})$,
$u_0=0.2\,\xi\,(\simeq4\,{\rm\AA})$, we obtain, with the values of
the material constants estimated in Sec.~\ref{sec:oom}, the typical
energy scale $\pi B\,r_0\,\xi\,u_0^2\simeq10\,k_{\rm B}T$. The
boundary tilt-difference angle is then given by
$\alpha_0=x\,u_0\,\sqrt{B/t'}\simeq x\,u_0/\xi\simeq x\times10^\circ$
(for $\lambda\simeq t'$ as consistently assumed in Sec.~\ref{sec:oom}).
Therefore, in Fig.~\ref{fig:nrj.cne0}, the depths of the two minima
are $\simeq25\,k_{\rm B}T$ and $\simeq3\,k_{\rm B}T$, respectively. Two
distinct crystals might therefore appear: one with a distance between
the boundaries of the particles of
$2(R-r_0)\simeq2\times1.2\,\xi\simeq50\,{\rm\AA}$, the other with a
much larger separation
$2(R-r_0)\simeq2\times4.5\,\xi\simeq180\,{\rm\AA}$~\cite{par_surface}.

\begin{figure}
  \resizebox{0.5\textwidth}{!}
  {\hspace{30pt}\includegraphics{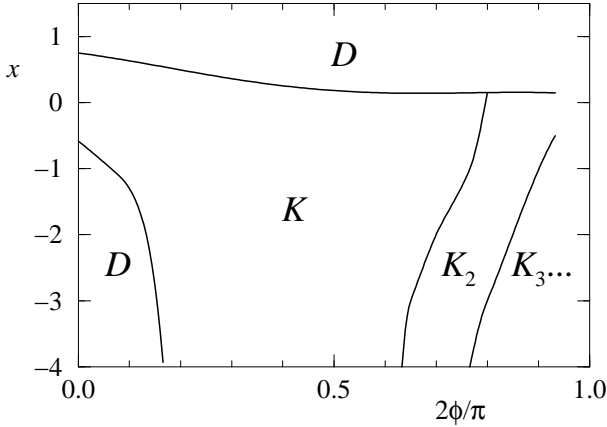}\hspace{30pt}}
\caption{
  Phase diagram for a membrane with $\xi\simeq20\,{\rm\AA}$ containing
  dilation-tilt-difference inducing inclusions with radius
  $r_0=3\,\xi_\alpha(\simeq60\,{\rm\AA})$. (D) disordered phase. (K)
  crystal phase, (K$_n$) region where $n$ possible distinct
  crystalline phases with different particles separations are
  possible.}
\label{fig:Ks}
\end{figure}

If we consider the inclusions radius $r_0$ as fixed, the interaction
potential as a function of $R$ depends only on the parameters
$x$ and $\phi$, as can be seen from~(\ref{eq:nor2}).
Figure~\ref{fig:Ks} shows a phase diagram in the $(x,\phi)$ plane for
a collection of identical inclusions. The symbol (D) indicates a
disordered (D) gaseous state, (K) a crystal phase, and (K$_n$) the
possibility of $n$ distinct crystalline phases with different
separation distances. Again, the criterion for any crystal phase is an
energy minimum depth larger than $k_{\rm B}T$.

An interesting feature of the phase diagram of Fig.~\ref{fig:Ks} is
the asymmetry with respect to the change $x\!\leftrightarrow\!-x$
introduced by the dilation--tilt-difference coupling: crystal phases
are more likely to occur for $x\!<\!0$, i.e., for a {\em
  thick-convex} inclusion ($u_0\!>\!0$ and $\alpha_0\!<\!0$) or for a
{\em thin-concave} inclusion ($u_0\!<\!0$ and $\alpha_0\!>\!0$). This
symmetry-breaking follows from the sign $c\!>\!0$ of the
dilation--tilt-difference coupling, that we have assumed, in order to
favor the situation depicted in Fig.~\ref{fig:coupling}.

\begin{figure}
  \resizebox{0.5\textwidth}{!}
  {\hspace{30pt}\includegraphics{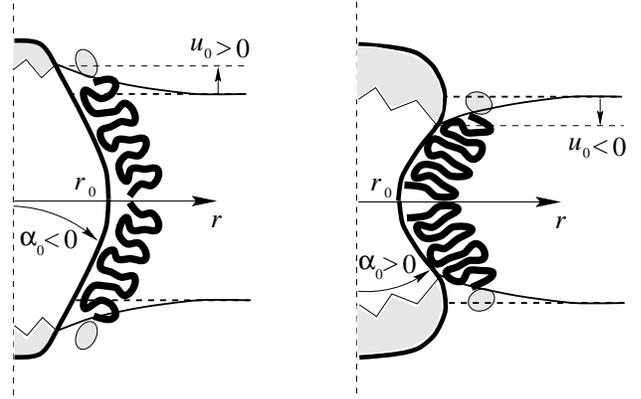}\hspace{30pt}}
\caption{
  Sketch of the inclusions that tend to form $2D$ crystals ($x\pp0$).
  The dashed lines show the monolayers of the unperturbed membrane.
  (Left) Thick-convex inclusion ($u_0\!>\!0$ and $\alpha_0\!<\!0$).
  (Right) Thin-concave inclusion ($u_0\!<\!0$ and $\alpha_0\!>\!0$).}
\label{fig:crfyl}
\end{figure}

\begin{figure}
  \resizebox{0.5\textwidth}{!}
  {\hspace{30pt}\includegraphics{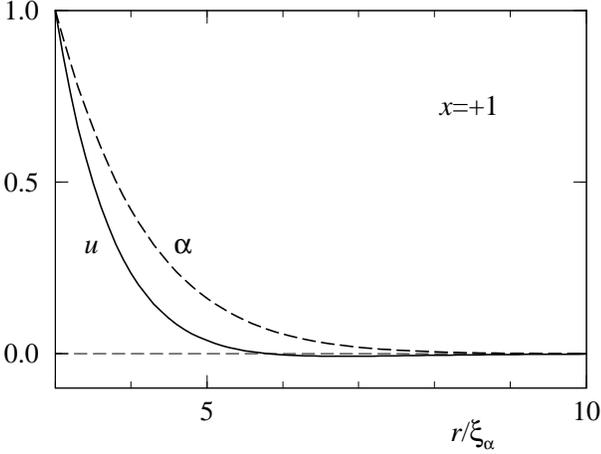}\hspace{30pt}}
\caption{
  Dilation and tilt-difference distortions around an isolated
  inclusion with $x\pg0$. The inclusion radius is
  $r_0=3\,\xi(\simeq60\,{\rm\AA})$ and the coupling corresponds to
  $\phi\!=\!0.2\times\pi/2$. Due to the latter, the tilt-difference
  tail expands (dashed line) and the dilation tail retracts (solid
  line), thereby favoring repulsion, i.e., a disordered phase.}
\label{fig:KDplus}
\end{figure}

This asymmetry can be explained by simple arguments. First, let us
recall that if a non-zero boundary tilt-difference $\alpha_0$ is
present, the interaction is always repulsive at short distances.
Indeed, the tilt-difference must go from $\alpha_0$ to $-\alpha_0$
from one inclusion to the other. Conversely, the distortion associated
with the membrane dilation is attractive since the thickness mismatch
is the same on identical inclusions. Two situations are therefore
possible: if the dilation relaxes on a longer range than the
tilt-difference, a crystal phase can occur since there is a long-range
attraction followed by a short-range repulsion, whereas conversely the
repulsion simply dominates. We therefore have to understand how the
coupling affects the relative range of the dilation and
tilt-difference distortions. To simplify, let us rewrite schematically
the interaction energy~(\ref{umc}) as
\begin{equation}
{\cal H}_{u\widehat m}\sim
  u^2+\xi^2\dot u^2+\dot u\alpha+\alpha^2+\xi^2\dot\alpha^2\,,
\end{equation}
Let us first assume $u_0,\alpha_0\!>\!0$, which corresponds to
$x\!>\!0$. To relax the positive dilation $u_0$, the membrane will set
$\dot u\pp0$. The term $\dot u\alpha$ being then negative, it reduces
the cost of making a gradient of $u$. Therefore the $u$ distortion
will relax on distance that is somewhat {\em shorter} than $\xi$: the
attractive dilation tail retracts (see Fig.~\ref{fig:KDplus}). From
the point of view of the tilt-difference, since $\dot u\pp0$, the
coupling $\dot u\alpha$ makes it as if the potential was of the type
$(\alpha-\alpha_{\rm m})^2$ with $\alpha_{\rm m}\pg0$.  Thus, on the
distance $\xi$, the tilt-difference relaxes only up to $\alpha_{\rm
  m}$; it therefore needs a {\em longer} distance to reach zero: the
repulsive tilt-difference tail expands (see Fig.~\ref{fig:KDplus}).
Then, for $x\pg0$, a disordered phase is more favored, since the
repulsive tail dominates at large distances.

\begin{figure}
  \resizebox{0.5\textwidth}{!}
  {\hspace{30pt}\includegraphics{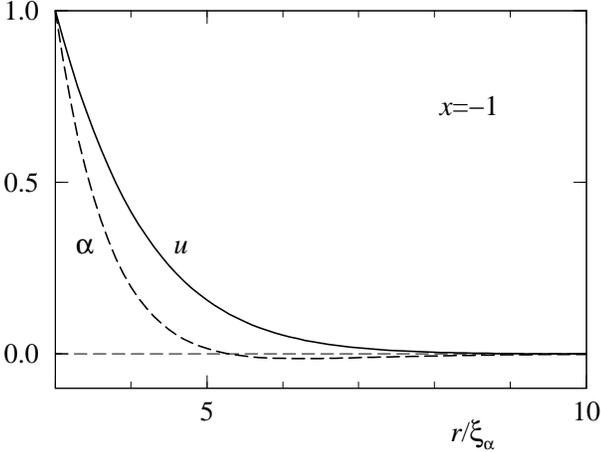}\hspace{30pt}}
\caption{
  Same as Fig.~\ref{fig:KDplus} but for $x\pp0$. Now the
  tilt-difference tail retracts and the dilation tail expands, thereby
  favoring attraction, i.e., a crystal phase.}
\label{fig:KDmoins}
\end{figure}

With still $u_0\pg0$, let us now assume $\alpha_0\pp0$, corresponding
to $x\pp0$. Now the term $\dot u\alpha$ is positive: building a
gradient of $u$ is more costly and therefore the attractive dilation
tail expands (see Fig.~\ref{fig:KDmoins}). The tilt-difference,
however, still experiences a potential of the type
$(\alpha-\alpha_{\rm m})^2$ with $\alpha_{\rm m}\pg0$, but it now
starts from a negative value $\alpha_0$. On a distance $\xi$ it would
reach the equilibrium value $\alpha_{\rm m}\pg0$, it therefore reaches
zero on a distance now shorter that $\xi$: the repulsive
tilt-difference tail retracts (cf.  Fig.~\ref{fig:KDmoins}).  Thus,
for $x\pp0$, a crystal phase is more favored, since the attractive
tail dominates at large distances (and then the repulsive one at short
distances).

\subsection{Shape-tilt induced interactions in an array of inclusions}

We now focus on the shape ($h$) and tilt (${\bf m}$) distortion modes
induced by the inclusions. Assuming again revolution symmetry in the
Wigner-Seitz cell,
\begin{equation}
  h=h(r) \quad{\rm and}\quad {\bf m}=\theta(r)\,{\bf
  e}_{r}\,,
\end{equation}
the most general solution of the equilibrium
equations~(\ref{hta}-\ref{htb}) takes the form
\begin{eqnarray}
  h&=&
(ar^2\!+\!b)\log{r}\!+\!cr^2\!+\!d\!+\!
A\,{\rm I}_0(qr)\!+\!
  B\,{\rm K}_0(qr)\,,\\
  \theta&=&-4\frac{L^2a}{\mu\,r}\!+\!
  \frac{qA}{\mu}\,{\rm I}_1(qr)-
  \frac{qB}{\mu}\,{\rm K}_1(qr)\,,
\end{eqnarray}
with
\begin{eqnarray}
  \mu&=&\frac{\gamma}{\kappa}\,,\\
  L&=&\frac{\gamma}{\sqrt{t\kappa}}\,,\\
  \xi_\theta&=&\sqrt{\frac{K_1}{t}}\,,\\
  q^{-1}&=&\sqrt{\xi_\theta^2-L^2}\,.
\end{eqnarray}
We assume $L<\xi_\theta$, i.e., $\gamma^2>K\kappa$, otherwise the
membrane undergoes the ripple instability of the $P_{\beta'}$ phase,
in which undulations and periodic tilt distortions
occur~\cite{ripple1,ripple2}. We assume also $L>0$, i.e., $\gamma>0$, since we
expect the molecules to tilt in such a way as to {\em relax\/} the
splay of the molecules in a curved membrane.

\begin{figure}
  \resizebox{0.5\textwidth}{!}
  {\hspace{10pt}\includegraphics{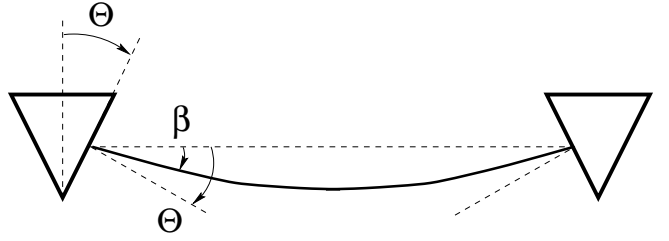}\hspace{10pt}}
\caption{
  Sketch of the membrane mid-surface shape ($h$) between conical
  inclusions.  The tilt of the lipid molecules at the inclusion
  boundary is $\beta-\Theta$.}
\label{fig:relax}
\end{figure}

To simplify, let us assume strictly $K_1=\kappa$. We then have
$L=\mu\,\xi_\theta$. Hence $0<\mu<1$ is now the only parameter
controlling the shape-tilt coupling. The six real unknowns $a$, $b$,
$c$, $d$, $A$, and $B$ are determined, according to the general
boundary conditions~(\ref{bch}-\ref{bct}) for an inclusion
with average cone angle $\Theta$, by
\begin{eqnarray}
  h|_{r_0}&=&h_0\,,\\
  \dot h|_{r_0}&=&\beta\,,\\
  \label{trade}
  \theta|_{r_0}&=&\beta-\Theta\,,\\
  h|_{R}&=&h_0\,,\\
  \dot h|_{R}&=&0\,,\\
  \theta|_{R}&=&0\,.
\end{eqnarray}
\begin{figure}
  \resizebox{0.5\textwidth}{!}
  {\hspace{30pt}\includegraphics{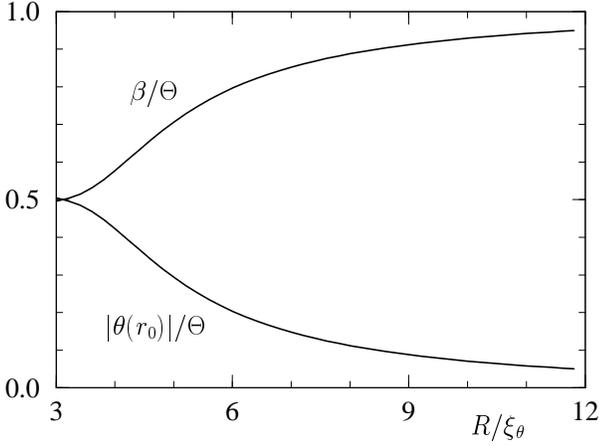}\hspace{30pt}}
\caption{Boundary tilt $\theta(r_0)$ and boundary membrane 
  inclination $\beta$ as a function of the distance $R$ between the
  inclusion. The inclusions radius is
  $r_0=3\,\xi_\theta(\simeq60\,{\rm\AA})$ and the tilt-shape coupling
  $\gamma$ is zero.}
\label{fig:angles}
\end{figure}
\begin{figure}
  \resizebox{0.5\textwidth}{!}
  {\hspace{30pt}\includegraphics{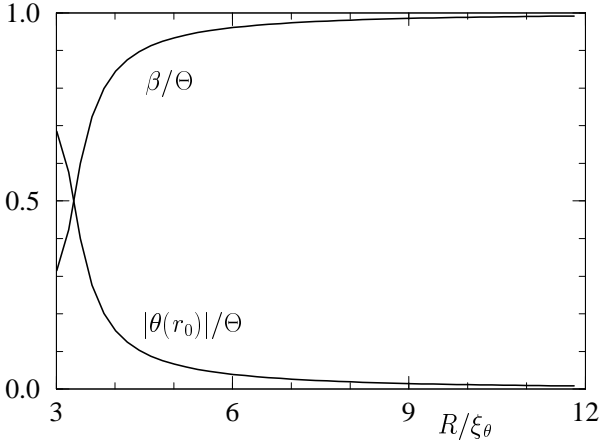}\hspace{30pt}}
\caption{Same as Fig.~\ref{fig:angles}, however 
  in the presence of a strong shape-tilt coupling corresponding to
  $\mu=0.9$.}
\label{fig:angles2}
\end{figure}
The latter three conditions are required by symmetry on the
Wigner-Seitz circle, at which the origin of the membrane height has
been chosen. After solving this system, the total membrane free energy
has to be minimized with respect to the free parameters $h_0$ and
$\beta$.  Assuming revolution symmetry, the general distortion
energy~(\ref{calH}) within the Wigner-Seitz cell takes the form
\begin{eqnarray}
  {\cal H}_{hm}=\pi&&\!\!\left[
    \kappa(r\,{\ddot h}+{\dot h}){\dot h}
    -\kappa\,h(r\,{\dot{\ddot h}}+{\ddot h}-{\dot h}/r)\right.\nonumber\\
  &&-\left.\gamma(r\,{\dot\theta}+\theta){\dot h}
    +\gamma\,h(r\,{\ddot\theta}+{\dot\theta}-\theta/r)\right.\nonumber\\
  &&-\left.\gamma(r\,{\ddot h}+{\dot h})\theta+
    K_1\,r\,\theta\,{\dot\theta}+K_1\,\theta^2
    \right]_{r_0}^R\,,
\end{eqnarray} 
where all the terms taken in $r\eg R$ vanish due to the boundary
conditions. The interaction has the following scaling property:
\begin{equation}\label{hhmnor}
  \frac{{\cal H}_{hm}}{\pi\kappa\,\Theta^2}=\overline{\cal H}_{hm}\!\left(
    \mu,\frac{r_0}{\xi_\theta},\frac{R}{\xi_\theta}\right)\,.
\end{equation}

The results are the following. Even in the absence of a shape-tilt
coupling, there is a trade between the shape and the tilt modes, which
is due to the boundary condition~(\ref{trade}). The membrane tends to
develop a tilt close to the inclusions in order to flatten its
shape. The typical solution for the membrane shape resembles that
sketched in Fig.~\ref{fig:relax}. The boundary tilt relaxes on a
distance $\simeq4\,q^{-1}$, typically of order a few $\xi_\theta$'s
unless $\mu$ is close to $1$.  The boundary tilt is a function of the
separation $R$ between the inclusions. When $R\gg\xi_\theta$, the
amplitude of the boundary tilt $\theta(r_0)$ is negligible: the
membrane curvature, which is small, only exerts a weak torque on the
tilt. Conversely, when the inclusions are close to contact, the
boundary tilt is a finite fraction of the inclusions average cone
angle $\Theta$. For $K_1=\kappa$, as we have assumed, this fraction is
exactly $1/2$ for $\mu=0$ (Fig.~\ref{fig:angles}).

\begin{figure}
  \resizebox{0.5\textwidth}{!}
  {\hspace{30pt}\includegraphics{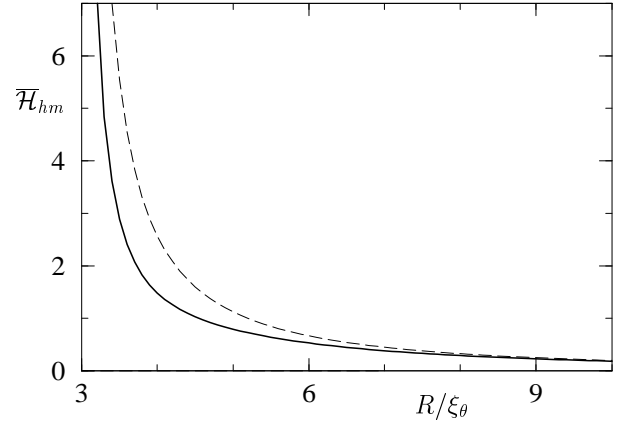}\hspace{30pt}}
\caption{Normalized interaction energy per
  inclusion $\overline{\cal H}_{hm}$ {\em vs.}~inclusions separation
  $R$. The inclusions radius is
  $r_0=3\,\xi_\theta(\simeq60\,{\rm\AA})$ and the tilt-shape coupling
  $\gamma=0$. The dashed curve corresponds to the case where the tilt
  is not allowed.}
\label{fig:nrjtilt}
\end{figure}

In the presence of a strong shape-tilt coupling ($\mu$ close to $1$),
the tilt relaxes on a distance $q^{-1}$ significantly shorter than
$\xi_\theta$, and the boundary tilt $\theta(r_0)$, which gets somewhat
larger at contact, actually relaxes more rapidly with the distance
between the inclusions (Fig.~\ref{fig:angles2}). Thus, except when the
inclusions are very close to one another, the tilt is rapidly
negligible.  The reason is that the tilt set in order to flatten the
membrane is {\em always\/} costly form the point of view of the
$-\gamma\,\nabla^2 h\,(\nabla\cdot{\bf m})$ coupling (when
$\gamma>0$).  Hence the coupling does not favor an expansion of the
tilt distortion, while in the preceding section the
dilation--tilt-difference coupling did favor an expansion of the
tilt-difference for $x<0$, which produced spectacular effects.

Let us estimate the magnitude of the interaction energy, which is
given by the normalization factor $\pi\kappa\,\Theta^2$
in~(\ref{hhmnor}). With $\xi_\theta\simeq20\,{\rm\AA}$, a typical
protein size $r_0=3\,\xi_\theta(\simeq60\,{\rm\AA})$ and
$\Theta\simeq10^\circ$, we obtain
$\pi\kappa\,\Theta^2\simeq2.5\,k_{\rm B}T$. Figure~\ref{fig:nrjtilt}
shows the interaction energy per inclusion in the case of zero
shape-tilt coupling. The interaction is always repulsive; it diverges
at small separations as $(R-r_0)^{-1}$ and tends asymptotically towards
the exact form
\begin{equation}
{\cal H}_{h}=2\pi\kappa\,\Theta^2\,\frac{r_0^2}{R^2-r_0^2}\,,
\end{equation}
which can be calculated analytically by completely neglecting the
tilt. As it is apparent in Fig.~\ref{fig:nrjtilt}, the tilt relaxes
some of the interaction energy at short inclusions separations. For
$\mu\to1$, we find that ${\cal H}_{hm}\to{\cal H}_{h}$ at all
separations. The effect of the tilt is therefore negligible.

\section{Conclusions}

We have developed an elastic model for membranes that describes at the
same level large- and short-scale distortions of the bilayer.
Strictly speaking, such a continuum theory at a molecular scale should
not be expected to give more than semi-quantitative results (see
Sec.~\ref{sec:trunc}). Nevertheless our hope is that the theory
captures the qualitative trends of the competitions between the
different elastic variables.  Using a systematic expansion in the
monolayers profiles and tilts, we have shown that the average membrane
shape ($h$) is coupled to the average molecular tilt (${\bf m}$), both
being decoupled (at lowest order) from the membrane dilation ($u$) and
the difference in the monolayers tilts (${\bf\widehat m}$), which are
coupled together.

We have used this model to study the contribution of the membrane
elasticity to the short- and long-range interactions among inclusions.
Because the boundary conditions at a membrane inclusion are decoupled
in the same way as the elastic variables, the interaction energy can
be calculated as simply the sum of a dilation--tilt-difference
contribution ($u$--${\bf\widehat m}$) and a shape-tilt contribution
($h$--${\bf m}$).

Membrane inclusions generally have a slightly convex or concave
hydrophobic core of thickness different from that of the bilayer. Such
inclusions will excite the coupled dilation--tilt-difference
($u$--${\bf\widehat m}$) mode. The thickness mismatch creates an
energetic dilation corona around the inclusions and yields an {\em
  attraction} between like inclusions: no extra distortion occurs when
the coronas overlap since the boundary dilations match. The
tilt-difference, however, yields a {\em repulsion} between like
inclusions: going from $\alpha_0$ to $-\alpha_0$, it develops a strong
gradient when the coronas overlap.  Inclusion producing no
tilt-difference aggregate, while inclusions producing a nonzero
tilt-difference either repel one another or favor $2D$ crystals. The
latter situation arises for small tilt-differences, or when the
dilation corona extends further than the tilt-difference corona.

When the dilation--tilt-difference coupling is large, the distortions
in the coronas exhibit damped oscillations. This effect occurs because
of the vicinity of a ``ripple'' instability in which both the membrane
dilation and tilt-difference become unstable.  The inter-particle
potential develops then several minima, which implies the possible
coexistence of different crystals of inclusions having different
lattice spacings. The latters can be significantly larger than the
inclusions size. The inclusions most likely to form $2D$ crystals are
those with either a {\em long-convex} or a {\em short-concave}
hydrophobic core, i.e., those disfavored from the point of view of the
$c\,\nabla u\cdot{\bf\widehat m}$ coupling. This is because the
gradient of $u$ being more costly, the dilation corona extends
(favoring ``long-range'' attraction), while at the same time the
dilation corona shrinks (making the repulsion occur only at smaller
separations).  Conversely, short-convex and long-concave inclusions
have a dominant repulsion and should form disordered phases.

Membrane inclusions generally have also a slightly conical shape.
Hence they excite the coupled shape-tilt ($h$--${\bf m}$) mode. In
first approximation, the conical shape constrains the membrane to
depart with a contact angle $\Theta$ relative to the inclusion axis.
The energy stored in the curvature of the membrane yields a repulsion
between like inclusions in an array that diverges at short distances
as $R^{-1}$ and fall off as $R^{-2}$. This is a many body effect,
since the interaction between a pair of inclusions falls off more
rapidly, as $R^{-4}$~\cite{goulian}. In the latter case, the
inclusions axes rotate away from one another in order to minimize the
curvature energy of the membrane. In an array of inclusions this
rotation is zero by symmetry.

If we allow for a tilt of the lipids, the membrane can depart with a
smaller contact angle $\beta$. In order to remain parallel to the
inclusions boundaries, the lipids tilt then by $\beta-\Theta$. When
the inclusions are far apart, the tilt is completely negligible since
the torque exerted by the membrane curvature on the tilt is weak.
Conversely, when the inclusions are distant a few times the membrane
thickness, the tilt becomes a finite fraction of $\Theta$. The
interaction energy is then reduced, however there is no qualitative
change in the interaction potential.  As for the shape-tilt coupling
$-\gamma\,\nabla^2 h\,(\nabla\cdot{\bf m})$, it reduces the relaxation
length of the tilt and simply reduces its effects. The reason is that
the tilt set in order to flatten the membrane is always costly form
the point of view of the coupling, for the expected positive sign of
$\gamma$. Hence the tilt does not propagate far away of the inclusions
in the vicinity of the ripple instability (where both the shape and
tilt modes become unstable).

\vspace{12pt}
{\bf Acknowledgments}
\vspace{12pt}

Useful discussions with A. Ajdari, P. Pincus and L. Peliti are
gratefully acknowledged. This work was partially supported by the NSF
Grants No. MRL DMR 91-23048 and 96-24091, and by the CNRS.



\begin{thebibliography}{99}
\bibitem{note} Note that in~\cite{dan2} the $(\nabla u)^2$ term would
  be restored if the lowest order term of the $u$-expansion would be
  considered a function $f_0(\Sigma,(\nabla\Sigma)^2)$ of the area per
  molecule {\em and its gradient}.
\bibitem{par_surface} In principle one should consider the free
  energy per surface unit ${\cal H}/R^2$, however this does not change
  significatively the position of the energy minimum.
\end{thebibliography}
\end{document}